# Quantum-Confined Tunable Ferromagnetism on the Surface of a van der Waals Antiferromagnet NaCrTe$_2$

Yidian Li[1], Xian Du[1], Junjie Wang[2,3], Runzhe Xu[1], Wenxuan Zhao[1], Kaiyi Zhai[1], Jieyi Liu[4], Houke Chen[4], Yiheng Yang[4], Nicolas C. Plumb[5], Sailong Ju[5], Ming Shi[5], Zhongkai Liu[6,7], Jiangang Guo[2], Xiaolong Chen[2], Yulin Chen[4,6,7\*], and Lexian Yang[1,8,9\*]

[1] *State Key Laboratory of Low Dimensional Quantum Physics, Department of Physics, Tsinghua University, Beijing 100084, China.*

[2] *Lab for Advanced Materials and Electron Microscopy, Institute of Physics, Chinese Academy of Sciences, Beijing 100083, China*

[3] *University of Chinese Academy of Sciences, Beijing 100049, China.*

[4] *Department of Physics, Clarendon Laboratory, University of Oxford, Parks Road, Oxford OX1 3PU, UK.*

[5] *Photon Science Division, Paul Scherrer Institut, CH-5232, Villigen, PSI, Switzerland*

[6] *School of Physical Science and Technology, ShanghaiTech University, Shanghai 201210, China.*

[7] *ShanghaiTech Laboratory for Topological Physics, Shanghai 200031, China.*

[8] *Frontier Science Center for Quantum Information, Beijing 100084, China.*

[9] *Collaborative Innovation Center of Quantum Matter, Beijing 100084, China.*

*e-mail: YLC: yulin.chen@physics.ox.ac.uk; LXY: lxyang@tsinghua.edu.cn*

**Abstract**

The surface of three-dimensional materials provides an ideal and versatile platform to explore quantum-confined physics. Here, we systematically investigate the electronic structure of Na-intercalated $CrTe_2$, a van der Waals antiferromagnet, using angle-resolved photoemission spectroscopy and *ab-initio* calculations. The measured band structure deviates from the calculation of bulk $NaCrTe_2$ but agrees with that of ferromagnetic monolayer $CrTe_2$. Consistently, we observe an unexpected exchange splitting of the band dispersions, persisting well above the Néel temperature of bulk $NaCrTe_2$. We argue that $NaCrTe_2$ features a quantum-confined 2D ferromagnetic state in the topmost surface layer due to strong ferromagnetic correlation in the $CrTe_2$ layer. Moreover, the exchange splitting and the critical temperature can be controlled by surface doping of alkali-metal atoms, suggesting a feasible tunability of the surface ferromagnetism. Our work not only presents a simple platform to explore tunable 2D ferromagnetism but also provides important insights into the quantum-confined low-dimensional magnetic states.

Two-dimensional (2D) magnetism violating the Mermin-Wagner theorem [1] not only features great scientific significance but also promises application potentials in electronic and spintronic devices [2-6]. Upon searching for robust 2D magnetic materials with desirable tunability, many interesting systems have emerged with attractive properties, such as the intrinsic long-range ferromagnetism in $CrI_3$ and $Cr_2Ge_3Te_6$ [7, 8], room-temperature ferromagnetism in ultrathin films of different chalcogenides [9-11], and quantum anomalous Hall effect in 5-layer $MnBi_2Te_4$ [12]. Van der Waals (vdW) heterostructures were also fabricated to explore the interplay between magnetism, topology, superconductivity, and ferroelectrics [2, 4, 5, 13]. However, the requirement for high-quality thin films or heterostructures makes it challenging to further investigate 2D magnetic systems. Alternatively, the surface of 3D materials naturally provides a versatile playground for exploring quantum-confined 2D physics, including 2D magnetism [14-16].

Among various 2D magnetic platforms, transition metal chalcogenides with high magnetic transition temperatures and feasible tunability have invigorated tremendous research interests [9, 17, 18]. A prototypical example is the room-temperature ferromagnetism in ultrathin $1T$-$CrTe_2$ films [11, 19, 20]. They exhibit many intriguing magnetic properties, including large magnetic anisotropy at room temperature [11], thickness-tunable Curie temperature [21], and colossal anomalous Hall effect [22]. However, air-unstable $CrTe_2$ films can only be fabricated using molecular beam epitaxy or chemical vapor deposition, and the bulk $CrTe_2$ crystal is metastable, which limits further investigation of $CrTe_2$-based 2D magnetism. Some crucial properties of $CrTe_2$ films, such as the Curie temperature and ferromagnetic (FM) mechanism, remain elusive [11, 19, 21, 23-25].

On the other hand, the metastable CrTe$_2$ crystal can be stabilized by intercalating alkali-metal atoms, as exemplified by NaCrTe$_2$, an A-type antiferromagnet with Néel temperature $T_N$ = 106 K [26, 27]. It retains the in-plane ferromagnetism in each layer [27, 28], which provides an alternative route to explore the intriguing CrTe$_2$-based 2D magnetism on the sample surface. In addition, NaCrTe$_2$ itself exhibits a spin-flip transition into FM spin alignment under a magnetic field accompanied by a giant negative magnetoresistance effect and tunable perpendicular magnetic anisotropy [26, 27], promising important applications in magneto-electronic devices.

In this work, we systematically investigate the electronic structure of vdW antiferromagnet NaCrTe$_2$ using high-resolution angle-resolved photoemission spectroscopy (ARPES) and *ab-initio* calculations. The surface layer of NaCrTe$_2$ is an interesting analog to the monolayer CrTe$_2$ or Na-decorated CrTe$_2$. The measured band structure is different from the calculation of bulk NaCrTe$_2$ in the antiferromagnetic (AFM) state. On the contrary, the calculation of FM CrTe$_2$ monolayer well reproduces the experiments. Consistently, we observe an unexpected exchange splitting of energy bands that characterizes a quantum-confined surface FM state with a critical temperature well above $T_N$ of bulk NaCrTe$_2$. Moreover, the surface FM state can be effectively controlled by surface doping of alkali-metal atoms. The quantum-confined ferromagnetism on the AFM bulk NaCrTe$_2$ provides a simple platform to explore the rich physics and application potentials of 2D magnetism in CrTe$_2$-based systems.

NaCrTe$_2$ crystallizes into a layered structure with the space group *P*-3*m*1 (No. 164), as schematically shown in Figure 1a. The bondless Na atoms occupy the octahedral interstitial sites between $CrTe_2$ layers (Figure 1a) and expand the lattice constants of the $CrTe_2$ layers into the FM region in its magnetic phase diagram [29, 30]. Figure 1b shows the hexagonal Brillouin zone and its surface projection, with high-symmetry points indicated. With decreasing temperature from room temperature, the magnetic moment first follows a Curie-Weiss behavior, then shows a sudden reduction below 106 K (Figure 1c), indicating an AFM transition. In the AFM state, the magnetic moments of Cr atoms align in an A-type AFM form along the out-of-plane direction with intralayer FM order (Figure 1a). Figure 1d shows the constant-energy contours around $\bar{\Gamma}$ at different binding energies. We observe two hole pockets $α_1$ and $α_2$ on the Fermi surface and another two hole pockets $δ_1$ and $δ_2$ in the constant-energy contours below about -0.1 eV.

Figures 2a and 2b show the measured band dispersions along the $\bar{\Gamma}\bar{K}$ and $\bar{\Gamma}\bar{M}$ directions, respectively. We observe two hole bands $α_1$ and $α_2$ crossing $E_F$ with similar Fermi velocities. At higher binding energies, there exist two extra hole bands $δ_1$ and $δ_2$ (Figure 2a). To unravel the origin of the two sets of splitting-like energy bands, we perform *ab-initio* calculations of the band structure of bulk $NaCrTe_2$ in the AFM state. Nevertheless, the calculation clearly deviates from the measured band structure as compared in Figures 2a-2c. The calculation of $NaCrTe_2$ not only overestimates the Fermi momenta of the energy bands but also suggests a clear anisotropy of the band splitting along $\bar{\Gamma}\bar{K}$ and $\bar{\Gamma}\bar{M}$. Moreover, there are three hole bands between -0.2 and -0.6 eV in the calculation instead of two hole bands in the experiment.

It is interesting to note that the experimental results are similar to those of CrTe$_2$ films [11]. Indeed, the calculation of monolayer CrTe$_2$ well reproduces the experimental results (Figure 2d). Figures 2e-2h compare the experimental and calculated band structures in a large energy and momentum range (Supporting Information, Figure S1). Again, the experiments fit much better to the calculation of monolayer CrTe$_2$ in the FM state than that of bulk NaCrTe$_2$ in the AFM state. Considering the surface sensitivity of ARPES experiments, we argue that the ARPES spectra of NaCrTe$_2$ are dominated by the photoelectrons emitted from the surface layer of Na-decorated CrTe$_2$.

To further decipher the nature of the observed splitting-like band dispersions, we perform detailed temperature-dependent ARPES measurements. As shown in Figure 3a, with increasing temperature, the $\alpha_1$ and $\alpha_2$ bands gradually approach each other and finally merge into one band $\alpha$ at 142 K. By fitting the momentum distribution curves (MDCs) at $E_F$ (see Figure 3b and Supporting Information Figure S4 for details), we can track the temperature evolution of the band splitting. As shown in Figure 3c, the observed band splitting resembles the behavior of a typical FM exchange splitting. To quantify the FM transition temperature, we fit the band splitting by the Weiss mean-field model (see Supporting Information Figure S6 for the fit of the data to Landau's phase transition theory)[31]. For the $J = 3/2$ system as in CrTe$_2$ [26], the model gives

$$\frac{M(T)}{M_S} = B_J\left(\frac{3J}{J+1}\cdot\frac{M(T)}{M_S}\cdot\frac{T_C}{T}\right) = \frac{4}{3}\coth\left(\frac{12}{5}\frac{M(T)}{M_S}\frac{T_C}{T}\right) - \frac{1}{3}\coth\left(\frac{3}{5}\frac{M(T)}{M_S}\frac{T_C}{T}\right) \quad (1),$$

where $M_s$ is the saturated magnetization, and $B_J(x)$ is the Brillouin function. Using this model, we obtain a critical temperature of $T_c = 128 \pm 4$ K (Figure 3c), well above the bulk $T_N = 106$ K. Similar analysis applies for the energy distribution curves (EDCs) at $k_{\|}$ = 0.24 Å$^{-1}$ (see Figure 2d and Supporting Information, Figure S5), which gives $T_c$ = 125 ± 5 K (Figure 3e). It is worth mentioning that the splitting of $\delta_1$ and $\delta_2$ bands shows exactly the same behavior (Supporting Information, Figure S3).

The comparison between ARPES experiments and *ab-initio* calculations, together with the temperature evolution of the band structure, compellingly evidences a surface FM state. Considering the A-type AFM property of NaCrTe$_2$ and the previously reported room-temperature ferromagnetism in ultrathin CrTe$_2$ films [11, 27], we argue that the observed exchange splitting is due to the quantum-confined FM state on the topmost surface of the system. We should also note that the strong temperature dependence of the exchange splitting that fits to the mean-field model is indicative of an itinerant FM state on the sample surface[32-34].

Interestingly, the surface ferromagnetism can be feasibly controlled by surface doping of alkali-metal atoms. As shown in Figures 4a-4d, the band splitting sensitively responds to the surface doping of Rb atoms. With increasing doping level, the $\alpha_1$ and $\alpha_2$ bands shift towards higher binding energies and the exchange splitting reduces. After surface doping for about 335 s, the $\alpha_1$ and $\alpha_2$ bands submerge below $E_F$, indicating a surface-doping-induced Lifshitz transition. Finally, the exchange splitting vanishes as summarized in Figure 4e. It is noteworthy that the chemical potential can be effectively tuned for more than 500 meV, suggesting an efficient tunability. More importantly, after the topmost CrTe$_2$ layer is covered by the doped alkali-metal atoms, the band structure becomes in good agreement with the *ab-initio* calculation of non-magnetic NaCrTe$_2$ (Supporting Information, Figure S2), confirming the suppression of the surface ferromagnetism.

We emphasize that the $T_c$ of the surface ferromagnetism is also tunable by surface doping. As shown in Figure 4f, we first dope Rb atoms on the sample surface for 120 s to partially suppress the exchange splitting, then conduct temperature-dependent measurements to determine the $T_c$ at this doping level. The subsequent temperature-dependent measurements (Figures 4g-4i) show a similar temperature evolution of the exchange splitting as in Figure 3. But the splitting now disappears at about 108 K (Figure 4j), well below that in the sample without doping, suggesting the $T_c$ of the surface ferromagnetism is lowered. In principle, the tunability of the ferromagnetism can be established either by modifying the exchange interaction between magnetic moments mediated by carrier density or by changing the electronic band structure, particularly the density of states (DOS) near $E_F$, as represented by the Ruderman–Kittel–Kasuya–Yosida (RKKY) and Stoner mechanism, respectively.

In vdW magnets, the interlayer magnetic coupling is usually one or two orders of magnitude weaker than the intralayer coupling. This leaves the possibility for the intermediate states where the interlayer magnetic order is quenched while the intralayer magnetic order survives during magnetic transitions [26, 27]. However, it is unexpected to observe such a large FM exchange splitting on the surface of AFM materials by ARPES [16, 35-37]. Similar band splitting with much smaller energy scale was also observed on the surface of antiferromagnetic topological insulator MnBi$_2$Te$_4$, whose origin awaits further investigation [36, 38, 39]. In the 3D AFM rare-earth heavy fermion compound $EuRh_2Si_2$ [16], surface electrons are polarized by the magnetic proximity effect, forming itinerant surface ferromagnetism; In transition metal oxide $PbCoO_2$, the enhanced surface density of states drives the Stoner transition [40]. However, both of them depend on the surface termination with relatively low $T_c$. On the contrary, the observed ferromagnetism is inherited from the 2D magnetism of monolayer $CrTe_2$. Therefore, the $T_c$ of the quantum-confined surface ferromagnetism here is much higher. By doping or removing alkali-metal atoms on the sample surface, the $T_c$ of the surface FM state can in principle be tuned in a large temperature range, which provides an ideal and unique playground for further exploration. On the other hand, the asymmetric coverage of alkali-metal atoms on the top and bottom of the surface $CrTe_2$ layer alludes to a large Dzyaloshinskii-Moriya interaction (DMI) [41], which is beneficial for the non-centrosymmetric FM order [42, 43], which may play an important role in the observed surface FM state.

In summary, we observe an unusual exchange splitting on the surface of a vdW AFM material, Na-intercalated $CrTe_2$, by ARPES and *ab-initio* calculation. The $T_c$ of the quantum-confined 2D FM state is well above the bulk AFM $T_N$. Both the magnitude of the exchange splitting and the $T_c$ of the surface ferromagnetism can be effectively tuned by the surface doping of alkali-metal atoms. Our work provides not only a rare platform to explore the intra- and inter-layer magnetic interactions in vdW materials but also a feasible method to tune their magnetic properties.

**Methods**

High-quality NaCrTe$_2$ single crystals were synthesized using the self-flux method [27]. Synchrotron-based ARPES experiments were performed at beamline ULTRA at the Swiss Light Source (SLS) using a Scienta DA30L analyzer. The convolved energy and angular resolutions were set to 15 meV and 0.2°, respectively. Laser-based ARPES experiments were performed at Tsinghua University using a Scienta DA30 analyzer and a 7 eV laser source. The overall energy and angular resolutions were 3 meV and 0.2°, respectively. The samples were cleaved *in-situ* and measured under ultrahigh vacuum below 6 × 10$^{-11}$ mbar. First-principles band structure calculations were performed using QUANTUM ESPRESSO code package [44] with a plane-wave basis. The exchange-correlation energy was considered under Perdew-Burke-Ernzerhof (PBE) type generalized gradient approximation (GGA) [45] with spin-orbit coupling included. Hubbard $U$ = 4.0 eV was applied to describe the localized 3$d$ orbitals of Cr atoms [46, 47]. The cut-off energy for the plane-wave basis was set to 600 eV. A Γ-centered Monkhorst-Pack $k$-point mesh of 13×13×3 was adopted in the self-consistent calculations.

**ASSOCIATED CONTENT**

**Data Availability Statement**

The data sets that support the findings of this study are available from the corresponding author upon reasonable request.

**Supporting Information**

The Supporting Information is available free of charge at [URL to be inserted].

Photon energy-dependent measurements of NaCrTe$_2$, comparison between the experimental and calculated band structure of non-magnetic CrTe$_2$ and NaCrTe$_2$, temperature-dependent energy splitting of the $\delta_1$ and $\delta_2$ bands extracted from the energy distribution curves at the $\bar{\Gamma}$ point, fitting details of the temperature-dependent momentum distribution curves, fitting details of the temperature-dependent energy distribution curves, fit of the temperature evolution of the band splitting to Landau's phase transition theory, and second derivative and curvature analysis of the band dispersions.

**Author Contributions**

L.X.Y. conceived the scientific project. Y.D.L., J.Y.L., H.K.C., Y.H.Y. carried out ARPES measurements and data analyses with the assistance of X.D., R.Z.X., W.X.Z., K.Y.Z., N.C.P., S.L.J., M.S., Z.K.L., and Y.L.C. X.D. performed the *ab-initio* calculations. Single crystals were synthesized and characterized by J.J.W., J.G.G. and X.L.C. All authors contributed to the scientific planning and discussion.

**Notes**

Authors declare that they have no competing interests.

**ACKNOWLEDGMENTS**

This work is funded by the National Key R&D Program of China (No. 2022YFA1403100 and No. 2022YFA1403200), the Beijing Natural Science Foundation (Grant No. Z200005), the National Natural Science Foundation of China (Grants No. 12275148). L. X. Y. acknowledges the support from Tsinghua University Initiative Scientific Research Program. The experiments in PSI are conducted under the proposal ID: 20211981.

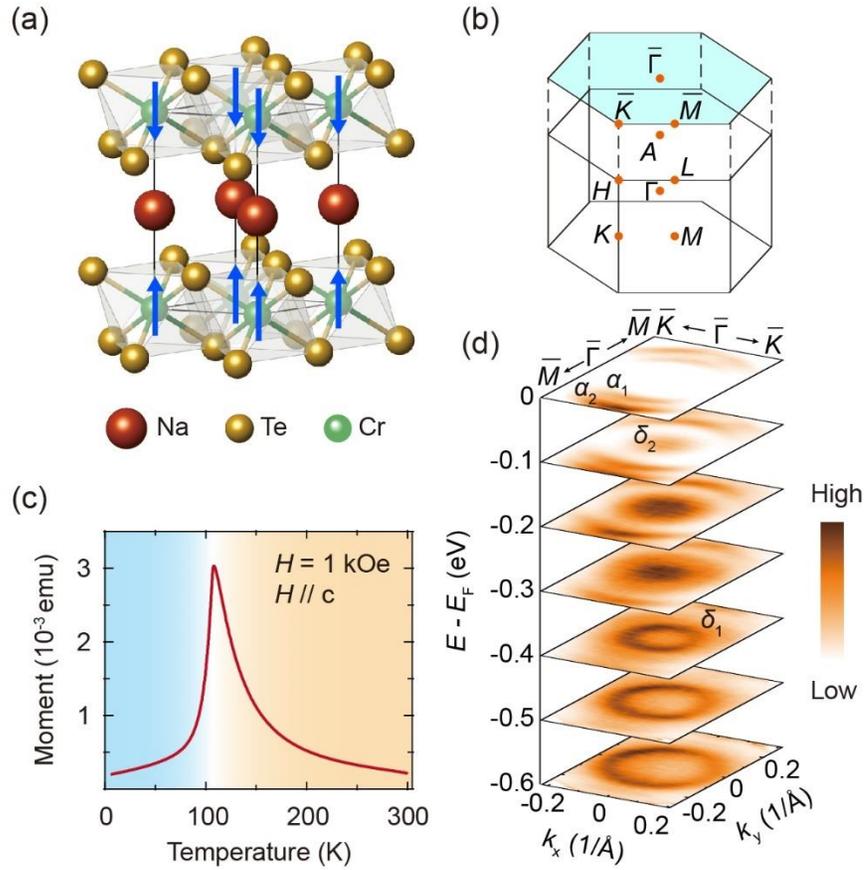

**Figure 1. Basic Properties of NaCrTe$_2$.** (a) Schematic of the crystal structure of NaCrTe$_2$. The blue arrows indicate the magnetic moment orientations of Cr atoms in the A-type antiferromagnetic (AFM) state. (b) Three-dimensional (3D) Brillouin zone (BZ) of NaCrTe$_2$ and its surface projection with high symmetry points indicated. (c) Field-cooled magnetization as a function of temperature showing the AFM transition at 106 K. (d) Constant-energy contours around $\bar{\Gamma}$ at selected binding energies.

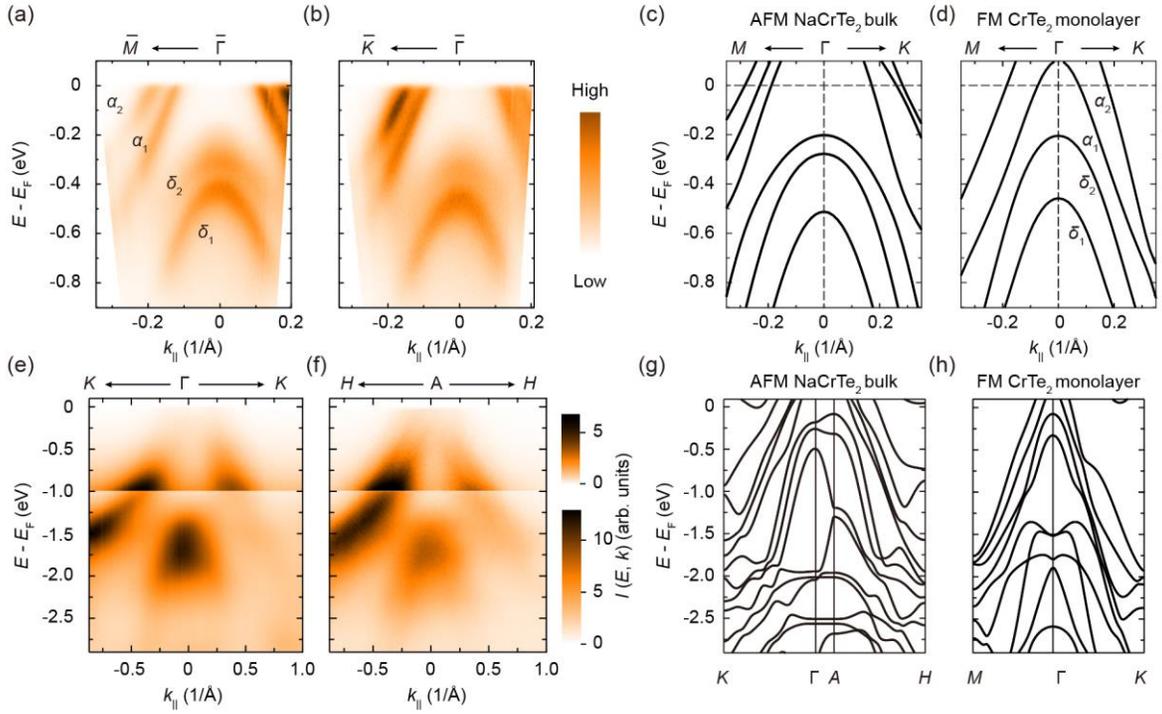

**Figure 2. Comparisons between experimental and calculated band structures.** (a,b) Band dispersions of NaCrTe$_2$ along the $\bar{\Gamma}\bar{M}$ (a) and $\bar{\Gamma}\bar{K}$ (b) directions. Data were collected using linear-horizontally (LH) polarized 7 eV laser at 80 K. (c) Calculated band structure of bulk NaCrTe$_2$ in the AFM state. (d) Calculated band structure of monolayer CrTe$_2$ in the ferromagnetic (FM) state. (e,f) Band dispersions of bulk NaCrTe$_2$ along the $AH$ and $\Gamma K$ directions. Data were collected using 75 eV and 100 eV photons with LH polarization at 20 K. The contrast of ARPES spectra above -1.0 eV was separately adjusted to enhance the states near the Fermi level ($E_F$). (g,h) Calculated band structures of bulk NaCrTe$_2$ in the AFM state (g) and monolayer CrTe$_2$ in the FM state (h).

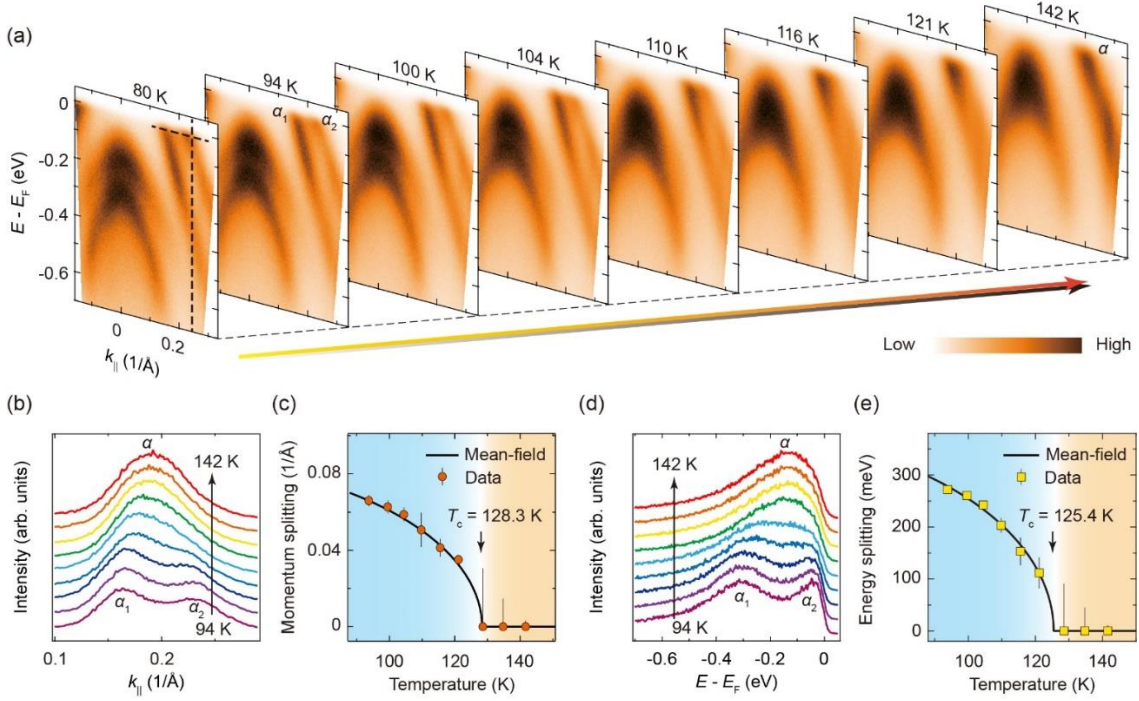

**Figure 3. Temperature dependence of the unusual exchange splitting.** (a) Temperature evolution of the band dispersion along the $\bar{\Gamma}\bar{M}$ direction. (b) Momentum distribution curves (MDCs) at selected temperatures showing the evolution of the exchange splitting. (c) Momentum splitting at $E_F$ [horizontal dashed line in the panel (a)] as a function of temperature. The black line is the fit of the data to the mean-field model. (d) Energy distribution curves (EDCs) at selected temperatures showing the evolution of the exchange splitting. (e) Energy splitting along the vertical dashed line in the panel (a) as a function of temperature. The black line is the fit of the data to the mean-field model.

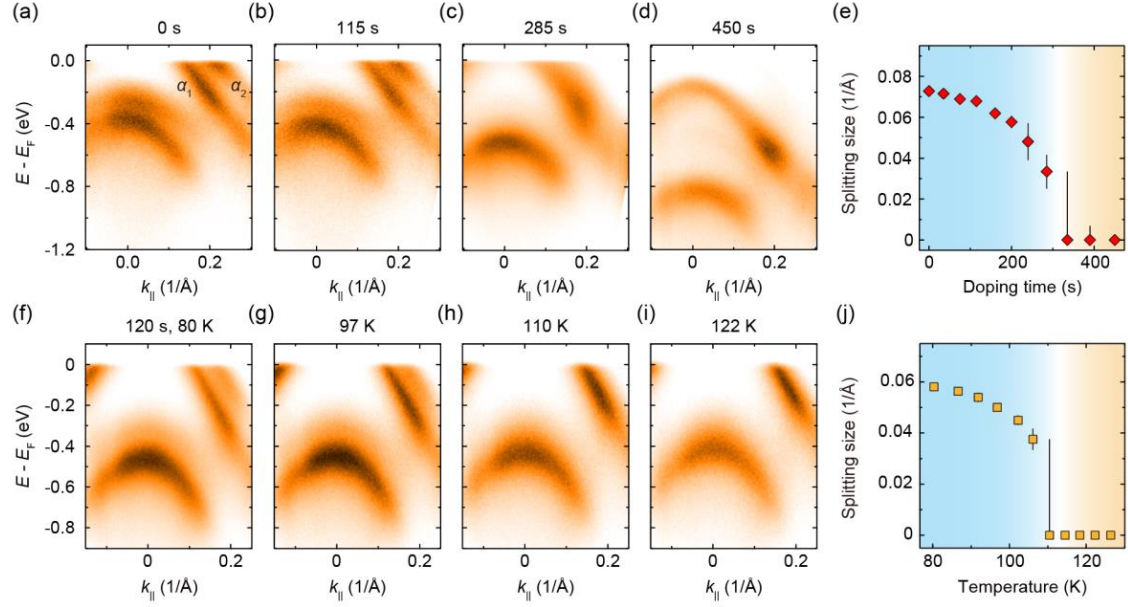

**Figure 4. Tunability of the surface ferromagnetism in NaCrTe$_2$ by alkali-metal atoms doping.** (a-d) Evolution of the band dispersion with surface doping time. The exchange splitting reduces and finally disappears at high doping levels. (e) Summary of the momentum splitting of the $α_1$ and $α_2$ bands at $E_F$ as a function of surface doping time. (f) Band structure of NaCrTe$_2$ with slight surface doping of Rb atoms. The exchange splitting is partially suppressed. (g-i) Band dispersions of the slightly doped NaCrTe$_2$ at selected temperatures. (j) Momentum splitting of the $α_1$ and $α_2$ bands in the slightly doped sample as a function of temperature.